\documentstyle[11pt,newpasp,twoside,epsf]{article}
\def\sax{\mbox{SAX J1808.4-3658}}
\def\be{\begin{equation}}
\def\ee{\end{equation}}

\def\lsim{\lower0.5ex\hbox{$\; \buildrel < \over \sim \;$}}

\def\ergcms{\mbox{erg\,cm$^{-2}$\,s$^{-1}$}}
\def\dm{\mbox{$\dot{M}$}}
\def\msun{\mbox{$M_{\odot}$}}
\def\ra{\mbox{$R_{\rm A}$}}
\def\rco{\mbox{$R_{\rm c}$}}
\def\ro{\mbox{$R_0$}}
\def\rs{\mbox{$R_{\rm s}$}}
\def\rms{\mbox{$R_{\rm ms}$}}
\def\dmmax{\mbox{$\dm_{\rm max}$}}
\def\dmmin{\mbox{$\dm_{\rm min}$}}
\def\fmax{\mbox{$F_{\rm max}$}}
\def\fmin{\mbox{$F_{\rm min}$}}
\def\etal{\mbox{\it et al.}}
\def\apj{Astrophys. J.\ }
\def\aap{Astron. Astrophys.\ }

\def\nat{\mbox{Nature\ }}
\def\sci{\mbox{Science\ }}

\def\edcomment#1{\iffalse\marginpar{\raggedright\sl#1\/}\else\relax\fi}
\reversemarginpar

\begin{document}
\title{DO STRANGE STARS EXIST IN THE UNIVERSE?}
%
%
\author{IGNAZIO BOMBACI}
\affil{Dipartimento di Fisica, Universit\`a di Pisa, and INFN Sez. di Pisa, 
      via Buonarroti, 2, I-56127 Pisa, Italy}

\begin{abstract}
Definitely, an affirmative answer to this question would have implications of 
fundamental importance for astrophysics (a new class of compact stars), 
and for the physics of strong interactions (deconfined phase of quark matter, 
and strange matter hypothesis). 
In the present work, we use observational data for the newly discovered  
millisecond X-ray  pulsar SAX J1808.4-3658 and for the atoll source 
4U~1728-34 to constrain the radius of the underlying compact stars. 
Comparing the mass--radius relation of these two compact stars 
with theoretical models for both neutron stars and strange stars, 
we argue that a strange star model is more consistent with  
SAX J1808.4-3658 and 4U~1728-34, and suggest that they are likely 
strange star candidates.  
\end{abstract}

\section{Introduction}
The possible existence of a new class of compact stars, which are made 
entirely of deconfined {\it u,d,s} quark matter ({\it strange quark  matter} 
(SQM)), is one of the most  intriguing aspects of modern astrophysics.  
These compact objects are called strange stars.  
They differ from neutron stars, where quarks are confined within neutrons, 
protons, and eventually within other hadrons ({\it e.g. hyperons}). 
The investigation of such a possibility is relevant not only for 
astrophysics, but for high energy physics too. 
In fact, the search for a deconfined phase of quark matter is one of the main 
goals in heavy ion physics. Experiments at Brookhaven National Lab's 
Relativistic Heavy Ion Collider (RHIC) and at CERN's Large Hadron 
Collider (LHC), will hopefully clarify this issue in the near future.  

The possibility that strange stars do exist is based on the so called 
{\it strange matter hypothesis}, formulated by Witten~(1984) (see also 
Bodmer, 1971).      
According to this hypothesis, strange quark matter, in equilibrium with 
respect to the weak interactions, could be the true ground state of strongly 
interacting matter rather than $^{56}Fe$, {\it i.e.} the energy per baryon 
of SQM must fulfil the inequality 
\be
\bigg( {{E}\over{A}} \bigg)_{SQM} \leq {{E(^{56}Fe)}\over{56}} \simeq 930~MeV, 
\label{eq:stableSQM}
\ee
at the baryon density where the pressure is equal to zero. 

If the strange matter hypothesis is true, then a nucleus with A nucleons, 
could in principle lower its energy by converting to a strangelet 
(a drop of SQM).  However, this process requires a very high-order 
simultaneous weak interactions to convert about a number A of {\it u} and 
{\it d} quarks of the nucleus into strange quarks. The probability for such 
a process is extremely low 
{\footnote{~It is proportional to $G_F^{2A}$, being $G_F$ the Fermi constant, 
and assuming a number $A$ of simultaneous weak processes.}}, 
and the mean life time for an atomic nucleus to decay to a strangelet is 
much higher than the age of the Universe. 
On the other hand, a step by step production of {\it s} quarks, at different 
times, will produce hyperons in the nucleus, {\it i.e.} a system (hypernucleus) 
with a higher energy per baryon with respect to the original nucleus.    
In addition, finite size effects (surface and shell effects) place a 
lower limit (A $\sim$ 10--100) on the baryon number of a stable 
strangelet even if bulk SQM is stable (Farhi \& Jaffe, 1984).   
Thus, according to the strange matter hypothesis, the ordinary state 
of matter, in which quarks are confined within hadrons, is a 
metastable state.    

 The success of traditional nuclear physics, in explaining an astonishing 
amount of experimental data, provides a clear indication that quarks in 
a nucleus are confined within protons and neutrons.  Thus, the energy per 
baryon $(E/A)_{ud}$ of {\it u,d} quark matter (nonstrange quark matter) 
must be higher than the energy per baryon of nuclei  
\be
\bigg( {{E}\over{A}} \bigg)_{ud} \geq  930~MeV + \Delta ,  
\label{eq:stableNSQM}
\ee
being $\Delta \sim 4$~MeV a quantity which accounts for the lower energy 
per baryon of a finite chunk ($A \sim 250$) of nonstrange quark matter with 
respect to the bulk ($A \rightarrow \infty$) case (Farhi \& Jaffe, 1984). 
These stability conditions (eq.s (1) and (2)) in turn may be used to 
constrain the parameters entering in models for the equation of state (EOS) 
of SQM. As we show below, the existence of strange stars is allowable 
within the uncertainties inherent in perturbative Quantum Chromo-Dynamics 
(QCD). Thus {\it strange stars may exist in the Universe}.     

\section{The equation of state for strange quark matter}
From a basic point of view the equation of state for SQM should be calculated 
solving the equations of QCD. As we know, such a fundamental approach is 
presently not doable. Therefore one has to rely on phenomenological models.  
In this work, we discuss two phenomenological models for the EOS 
of strange quark matter. 
The first one is a well known model related to the MIT bag model 
(Chodos \etal\ 1974) for hadrons. 
The second one is a new model developed by Dey \etal\ (1998). 

At very high density SQM behaves as a relativistic gas of weakly interacting 
fermions. This is a consequence of one of the basic features of QCD, namely 
asymptotic freedom.  
To begin with consider the case of massless quarks, and 
consider gluon exchange interactions to the first order in the  
QCD structure constant $\alpha_c$. Under these circumstances the EOS of 
$\beta$--stable SQM can be written in the parametrical form:   
\be 
   \varepsilon =  K n_B^{4/3} + B,  \qquad
     P   = {1\over{3}} K n_B^{4/3} - B,  ~\qquad
     K \equiv {9\over{4}} \pi^{2/3}  
        \bigg(1 + {{2\alpha_c}\over{3\pi}}\bigg) \hbar c 
\label{eq:eosB1}
\ee 
$\varepsilon$ being the energy density,  and  $P$ the pressure. 
Eliminating the baryon number density $n_B$ one gets:  
\be
                     P   = {1\over{3}} (\varepsilon - 4 B) 
\label{eq:eosBag}
\ee  
Here $B$ is a phenomenological parameter which represents the difference 
between the energy density of ``perturbative vacuum'' and true QCD vacuum.   
$B$ is related to the ``bag constant'' which in the MIT bag model for  
hadrons (Chodos \etal\ 1974) gives the confinement of quarks within   
the hadronic bag.    
The density of zero pressure SQM is just $\rho_s = 4B/c^2$. 
This is the value of the surface density of a bare strange star.  
Taking a non-vanishing value for the mass $m_s$ of the strange quark,  
the EOS becomes more involved (see {\it e.g.} Farhi \& Jaffe, 1984)  
with respect to the simple expression (4).  
However, for $m_s = 100$--300~MeV, equation (4) is less 
than 5\% different from the ``exact'' case for $m_s\neq 0$. 
In summary, in this model for the equation of state for SQM there are 
three phenomenological parameters, namely: $B$, $m_s$, and $\alpha_c$.  
It is possible to determine ranges in the values of these parameters in which 
SQM is stable, and nonstrange quark matter is not (Farhi \& Jaffe, 1984). 
For example, in the case of non--interacting quarks ($\alpha_c=0$) one has  
$B \simeq$ 57--91~MeV/fm$^3$ for $m_s = 0$, and 
$B \simeq$ 57--75~MeV/fm$^3$ for $m_s = 150$~MeV.   

The schematic model outlined above becomes less and less trustworthy 
going from very high density region (asymptotic freedom regime) to lower 
densities, where confinement (hadrons formation) takes place.  
Recently, Dey {\it et al.} (1998) derived a new EOS for SQM  using a 
``dynamical'' density-dependent approach to confinement.  
The EOS by Dey {\it et al.} has asymptotic freedom built in, shows 
confinement at zero baryon density, deconfinement at high density. 
In this model, the quark interaction is described by a colour-Debye-screened  
inter-quark vector potential originating from gluon exchange, 
and by a density-dependent scalar potential which restores chiral 
symmetry at high density (in the limit of massless quarks).    
The density-dependent scalar potential arises from the density dependence of 
the in-medium effective quark masses $M_q$, which, in the model by Dey \etal 
(1998), are taken to depend upon the baryon number density according to
\be 
M_q = m_q + 310 \cdot sech\bigg(\nu {{n_B}\over{n_0}}\bigg)
                                   \qquad  \qquad  ({\rm MeV}),    
\ee
where $n_0 = 0.16$~fm$^{-3}$ is the normal nuclear matter density,  
$q (= u,d,s)$ is the flavor index, and $\nu$ is a parameter.   
The effective quark mass $M_q(n_B)$ goes from its constituent masses 
at zero density, to its current mass $m_q$,  as $n_B$ goes to infinity. 
Here we consider two different parameterizations of the EOS by Dey 
{\it et al.}, which correspond to a different choice for the parameter $\nu$. 
The  equation of state SS1 (SS2) corresponds to $\nu = 0.333$ ($\nu = 0.286$). 
These two models for the EOS give absolutely stable SQM according to the 
strange matter hypothesis. 

\section{Strange star candidates}
To distinguish whether a compact star is a neutron star or 
a strange star, one has to find a clear observational signature.
There is a striking qualitative difference in the mass--radius (MR)   
relation of strange stars with respect to that of neutron stars (see Fig.~1). 
For strange stars with ``small'' ($M << M_{max}$) gravitational mass, 
$M$ is proportional to $R^3$. 
In contrast, neutron stars have radii that decrease with increasing mass.  
This is a consequence of the underlying interaction between the stellar 
constituents which makes ``low'' mass strange stars self-bound objects
(see {\it e.g.} Bombaci 1999) contrary to the case of neutron stars 
which are bound by gravity  
{\footnote{~As an idealized example, remember that pure neutron matter is 
not bound by nuclear forces.}}.
As we know, there is a minimum mass for a neutron star 
($M_{min} \sim 0.1~M_\odot$). In the case of a strange star, there is 
essentially no minimum mass. 
As the central density $\rho_c \to \rho_s$ (surface density), a 
strange star (or better a strangelet for very low baryon number) is a 
self--bound system, until the baryon number becomes so low that finite size 
effects destabilize it.   
%


\subsection{SAX J1808.4-3658}

The transient X-ray burst source \sax\ was discovered in September 1996  
by the BeppoSAX satellite.   
Two bright type-I X-ray bursts were detected, each lasting less than 
30 seconds. Analysis of the bursts in \sax\ indicates that it is 4~kpc 
distant and has a peak X-ray luminosity of $6\times 10^{36}~$erg/s in its 
bright state, and a X-ray luminosity lower than $10^{35}~$erg/s in 
quiescence (in't Zand 1998).  
The object is nearly certainly the same as the transient X-ray source 
detected with the Proportional Counter Array (PCA) on board the Rossi 
X-ray Timing Explorer (RXTE) (Marshall, 1998). 
Coherent pulsations at a period of 2.49 milliseconds were discovered 
(Wijnands \& van der Klis 1998). 
The star's surface dipolar magnetic moment was derived 
to be less than $10^{26}$~G~cm$^3$ from detection of X-ray pulsations at 
a luminosity of $10^{36}$~erg/s (Wijnands \& van der Klis 1998), 
consistent with the weak fields expected for type-I X-ray bursters and 
millisecond radio pulsars (MS PSRs) (Bhattacharya \& van den Heuvel 1991).   
The binary nature  of \sax\ was firmly established with the detection of a 
2 hour orbital period (Chakrabarty \& Morgan 1998) as well as with the 
optical identification of the companion star (Roche \etal\ 1998).      
\sax\ is the first pulsar to show both coherent pulsations in its persistent 
emission and X-ray bursts, and by far the fastest-rotating, lowest-field 
accretion-driven pulsar known.
It presents direct evidence for the evolutionary link between  low-mass 
X-ray binaries (LMXBs)  and MS PSRs.    
\sax\ is the only known LMXB with an MS PSR.  
 
A mass--radius (MR) relation for the compact star in \sax\  has been 
recently obtained by Li \etal\ (1999a)
{\footnote{~see also Burderi \& King (1998), Psaltis \& Chakrabarty (1999).}} 
using the following two requirements.  
({\it i}) Detection of X-ray pulsations requires that the inner radius $\ro$ 
of the accretion flow should be larger than the stellar radius $R$. 
In other words, the stellar magnetic field must be strong enough to 
disrupt the disk flow above the stellar surface. 
({\it ii}) The radius $\ro$ must be less than the so-called co-rotation 
radius $\rco$, {\it i.e.} the stellar magnetic field must be weak enough 
that accretion is not centrifugally inhibited: 
\be  
     \ro   \lsim  \rco = [GM P^2/(4\pi^2)]^{1/3}.  
\ee
Here $G$ is the gravitation constant, $M$ is the mass of the star, 
and $P$ is the pulse period. The inner disk radius $\ro$ is generally 
evaluated in terms of the Alfv\'en radius $\ra$, at which the magnetic   
and material stresses balance (Bhattacharya \& van den Heuvel 1991):    
$\ro=\xi\ra=\xi[B^2R^6/\dm(2GM)^{1/2}]^{2/7}$, where $B$ and $\dm$ 
are respectively the surface magnetic field and the mass accretion 
rate of the pulsar, and $\xi$ is a parameter of order of unity 
almost independent of $\dm$ (Li 1997, Burderi \& King 1998).     
Since X-ray pulsations in \sax\ were detected over a wide range of mass  
accretion rate (say, from $\dmmin$ to $\dmmax$), the two conditions 
({\it i}) and ({\it ii}) give $R\lsim \ro(\dmmax)< \ro(\dmmin)\lsim \rco$.  
Next, we assume that the mass accretion rate $\dm$ is proportional to the 
X-ray flux $F$ observed with RXTE. This is guaranteed by the fact that the 
X-ray spectrum of \sax\ was  remarkably stable and there was only slight 
increase in the pulse amplitude when the X-ray luminosity varied by a factor 
of $\sim 100$ during the 1998 April/May outburst (Gilfanov \etal\ 1998, 
Cui \etal\ 1998, Psaltis \& Chakrabarty 1999). Therefore, Li \etal\ (1999a)  
get the following upper limit of the stellar radius:    
$ R < (F_{min}/F_{max})^{2/7} \rco$, or 
\be  
      R < 27.5  \bigg({{F_{min}}\over{F_{max}}}\bigg)^{2/7}  
                 \bigg({{P}\over{2.49~ms}}\bigg)^{2/3} 
                 \bigg({{M}\over{M_\odot}}\bigg)^{1/3} ~{\rm km},  
\label{eq:MR-sax}
\ee
where $\fmax$ and $\fmin$ denote the X-ray fluxes measured 
during X-ray high- and low-state, respectively, $\msun$ is the 
solar mass. 
\begin{figure} 
\plotfiddle{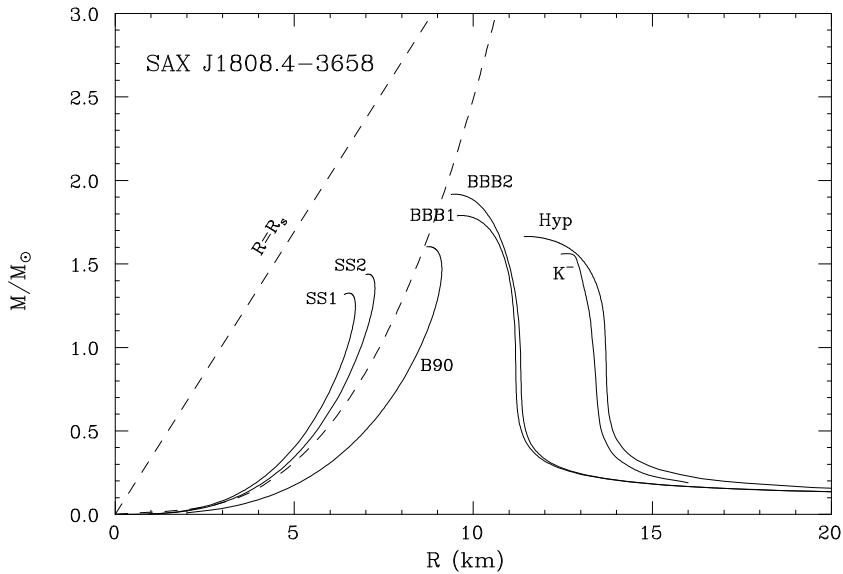}{7.0cm}{90}{50}{50}{200}{-60}
\caption{Comparison of the mass--radius relation of SAX J1808.4 -3658 
determined from RXTE observations with theoretical models of neutron 
stars and of strange stars. See text for more details.}   
\end{figure}
Note that in writing inequality (7) it is assumed that the pulsar's 
magnetic field is basically dipolar (see Li \etal\ 1999a for  
arguments to support this hypothesis)
{\footnote{~see also Psaltis \& Chakrabarty (1999) for a study of the 
influence on the MR relation for \sax\ of a quadrupole magnetic moment, 
and of a {\it non-standard} disk--magnetosphere interaction model.}}.  

Given the range of X-ray flux at which coherent pulsations were 
detected, inequality (7) defines a limiting curve in the 
mass--radius plane for SAX J1808.4-3658, as plotted in the dashed curve 
in Fig.~1. The authors of ref. (Li \etal\ 1999a) adopted the flux ratio 
$\fmax/\fmin\simeq 100$ from the observations that during the 1998 
April/May outburst, the maximum $2-30$ keV flux of \sax\ at the 
peak of the outburst was $\fmax\simeq 3\times 10^{-9}\,\ergcms$, 
while the pulse signal became barely detectable when the flux 
dropped below $\fmin\simeq 2\times 10^{-11}\,\ergcms$ 
(Cui \etal\ 1998, Psaltis \& Chakrabarty 1999).   
The dashed line $R = \rs \equiv 2GM/c^2$ represents the Schwartzschild 
radius  - the lower limit of the stellar radius to prevent the star collapsing 
into a black hole.  Thus the allowed range of the mass and radius of 
\sax\ is the region confined by these two dashed curves in Fig.~1. 

In the same figure, we report the theoretical MR relations (solid curves) 
for neutron stars given by some recent realistic models for the EOS of 
dense matter (see Li \etal\ 1999a for references to the EOS models). 
Models BBB1 and BBB2 are relative to ``conventional'' neutron stars 
({\it i.e} the core of the star is assumed to be composed by an 
uncharged mixture of neutrons, protons, electrons and muons in 
equilibrium with respect to the weak interaction).   
The curve labeled Hyp depicts the MR relation for a neutron 
star in which hyperons are considered in addition to nucleons as hadronic 
constituents. The MR curve labeled $K^-$ is relative to neutron stars  
with a Bose-Einstein condensate of negative kaons in their cores.   
It is clearly seen in Fig.~1 that none of the neutron star MR curves 
is consistent with \sax. Including rotational effects will shift 
the $MR$ curves to up-right in Fig.~1 (Datta \etal\ 1998), and does not 
help improve the consistency between the theoretical neutron star models 
and observations of \sax.   
Therefore \sax\ is not well described by a neutron star model. 
The curve B90 in Fig.~1 gives the MR relation for strange stars described 
by the schematic EOS (4) with B = 90 MeV/fm$^3$.  
The two curves SS1 and SS2  give the MR relation for strange stars calculated 
with the EOS by Dey et al. (1998). 
Figure~1 clearly demonstrates that a strange star model is more 
compatible with \sax\ than a neutron star one.  

\subsection{4U 1728-34}
Recently, Li et al. (1999b) investigated possible signatures for the 
existence of strange stars in connection with the newly discovered 
phenomenon of kilohertz quasi--periodic oscillations (kHz QPOs) in 
the X-ray flux from LMXB (for a review see van der Klis 2000).         
Initially,  kHz QPO data from various sources were interpreted assuming a 
simple {\it beat--frequency model} (see {\it e.g.} Kaaret \& Ford 1997). 
In many cases, two simultaneous kHz QPO peaks (``twin peaks'') are observed.  
The QPO frequencies vary and are strongly correlated with source flux.  
In the beat--frequency model the highest observed QPO frequency   
$\nu_u$ is interpreted as the Keplerian orbital frequency $\nu_K$  
at the inner edge of the accretion disk. 
The frequency $\nu_l$ of the lower QPO peak is instead interpreted as the 
beat frequency between $\nu_K$ and the neutron star spin frequency $\nu_0$, 
which within this model is equal to the separation frequency 
$\Delta\nu \equiv \nu_u - \nu_l$ of the two peaks. 
Thus $\Delta\nu$ is predicted to be constant.  
Nevertheless, novel observations for different 
kHz QPO sources have challenged this simple beat--frequency model.  
The most striking case is the source 4U 1728-34, where it was found that 
$\Delta\nu$ decreases significantly, from $349.3\pm1.7$ Hz to 
$278.7\pm11.6$ Hz, as the frequency of the lower kHz QPO increases   
(M\'endez \& van der Klis 1999). 
Furthermore, in the spectra observed by the RXTE for 4U 1728-34, 
Ford \& van der Klis (1998) found low-frequency Lorentian 
oscillations with frequencies between 10 and 50 Hz. 
These frequencies as well as the break frequency ($\nu_{break}$) of the 
power spectrum density for the same source were shown to be correlated with 
$\nu_u$ and $\nu_l$.  

A different model was recently developed by Osherovich \& Titarchuk (1999) 
(see also Titarchuk \& Osherovich 1999), who proposed a unified 
classification of kHz QPOs and the related observed low frequency 
phenomena.    
In this model, kHz QPOs are modeled as Keplerian oscillations under the 
influence of the Coriolis force in a rotating frame of reference 
(magnetosphere). The frequency $\nu_l$ of the lower kHz QPO peak is 
the Keplerian frequency at the outer edge of a viscous transition layer 
between the Keplerian disk and the surface of the compact star.   
The frequency $\nu_u$ is a hybrid frequency related to the 
rotational frequency $\nu_m$ of the star's magnetosphere by: 
$\nu_u^2=\nu_K^2+(2\nu_m)^2$.  
The observed low Lorentzian frequency in 4U 1728-34 is
suggested to be associated with radial oscillations in the viscous transition  
layer of the disk, whereas the observed break frequency is determined by 
the characteristic diffusion time of the inward motion of the matter in the 
accretion flow (Titarchuk \& Osherovich 1999). Predictions of this model 
regarding relations between the QPO frequencies mentioned above compare 
favorably with recent observations for 4U 1728-34, Sco X-1, 4U 1608-52, 
and 4U 1702-429.  

\begin{figure} 
\plotfiddle{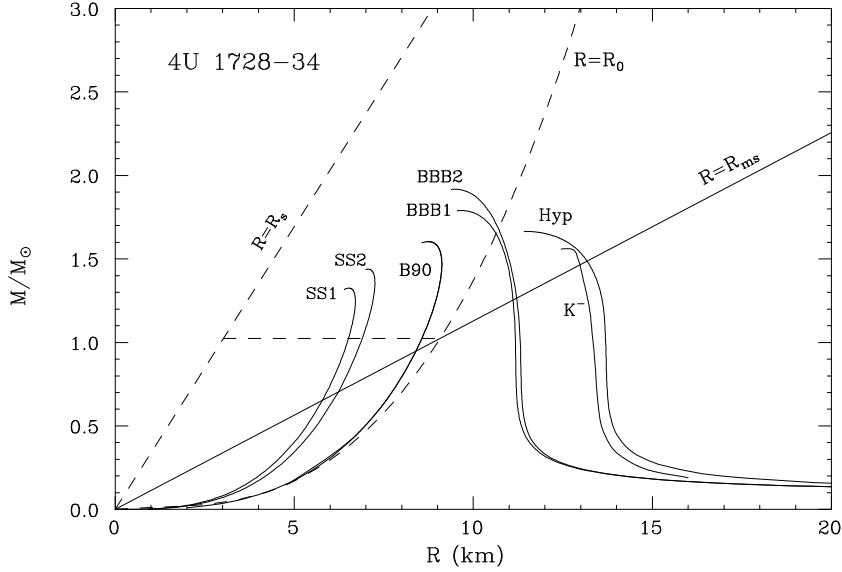}{7.0cm}{90}{50}{50}{200}{-60}
\caption{Comparison of the $MR$ relation of 4U 1728-34 determined from RXTE
observations with theoretical models of neutron stars and of strange
stars. The range of mass and radius of 4U 1728-34 is allowed in the
region outlined by the dashed curve $R=\ro$, the horizontal dashed line, 
and the dashed line $R=R_s$. The solid curves represents theoretical MR 
relations for neutron stars and strange stars.}  
\end{figure}

The presence of the break frequency and the correlated Lorentzian frequency 
suggests the introduction of a new scale in the phenomenon. 
One attractive feature of the model by Titarchuk \& Osherovich (1999)
is the introduction of such a scale in the model through the Reynolds number 
for the accretion flow.  
The best fit for the observed data was obtained by Titarchuk \& Osherovich 
(1999) when  
\begin{equation} 
a_k=(M/M_{\odot})(\ro/3\rs)^{3/2}(\nu_0/364\,{\rm Hz})=1.03,
\end{equation}
where $M$ is the stellar mass, 
$\ro$ is the inner edge of the accretion disk
{\footnote{~In the expression for $a_k$ reported in 
Titarchuk \& Osherovich (1999), one has $x_0 = \ro/\rs$, where $\ro$ is 
erroneously indicated as the neutron star radius 
(Titarchuk, private communication).}}, 
$\rs$ is the Schwarzschild radius, and $\nu_0$ is the spin
frequency of the star. 
Given the 364 Hz spin frequency of 4U 1728-34 (Strohmayer et al. 1996),  
the inner disk radius can be derived from the previous equation.  
Since the innermost radius of the disk must be larger than the radius $R$ 
of the star itself, this leads to a mass-dependent upper bound on the 
stellar radius,
\begin{equation}
  R \leq \ro  \simeq  8.86~ a_k^{2/3} (M/M_{\odot})^{1/3}\,{\rm km},
\end{equation}
which is plotted by dashed curve in Fig.~2. 

A second constraint on the mass and radius of 4U 1728-34 results from 
the requirement that the inner radius $\ro$ of the disk must be larger than
the radius of the last stable circular orbit $\rms$ around the star:  
\be 
                      \ro \geq \rms. 
\ee
To make our discussion more transparent, neglect for a moment the 
rotation of the compact star.  
For a non-rotating star $\rms = 3 \rs$, then the second condition gives:
\be 
   \ro \geq 3 \rs = 8.86~ \big(M/M_\odot\big)~{\rm km}. 
\ee
Therefore, the allowed range of the mass and radius for 
4U1728-34 is the region in the lower left corner of the MR plane confined 
by the dashed curve ($R=\ro$), by the horizontal dashed line, and by 
the Schwartzschild radius (dashed line $R=R_s$).  
In the same figure, we compare with the theoretical MR relations 
for non-rotating neutron stars and strange stars, for the same models 
for the EOS considered in Fig. 1.  
It is clear that a strange star model is more compatible with 4U 1728-34 
than a neutron star one. 
Including the effects of rotation ($\nu_0 = $364 Hz) in the calculation 
of the theoretical MR relations and $\rms$,  does not change the 
previous conclusion (Li \etal ~1999b).    

\section{Final remarks}
The main result of the present work ({\it i.e.} the likely existence 
of strange stars) is based on the analysis of observational data  
for the X-ray sources SAX J1808.4-3658 and 4U~1728-34.  
The interpretation of these data is done using {\it standard} models for 
the accretion mechanism, which is responsible for the observed phenomena.     
The present uncertainties in our knowledge of the accretion mechanism, 
and the disk--magnetosphere interaction,  do not allow us to definitely 
rule out the possibility of a neutron star for the two X-ray sources 
we discussed. 
For example, making {\it a priori} the {\it conservative} assumption that 
the compact object in \sax\ is a neutron star, and using a MR relation similar 
to our eq. (7) Psaltis \& Chakrabarty (1999) try to constrain 
disk--magnetosphere interaction models or to infer the presence of a 
quadrupole magnetic moment in the compact star. 

 \sax\ and 4U~1728-34 are not the only LMXBs which could harbour a 
strange star. Recent studies have shown that the compact objects  
associated with the X-ray burster 4U 1820-30 (Bombaci 1997), the bursting 
X-ray pulsar GRO J1744-28 (Cheng \etal\ 1998b) and the X-ray pulsar 
Her X-1 (Dey \etal\ 1998) are likely strange star candidates.  
For each of these X-ray sources (strange star candidates) the conservative 
assumption of a neutron star as the central accretor would require 
some particular (possibly {\it ad hoc}) assumption about the nature of the 
plasma accretion flow  and/or the structure of the stellar magnetic field. 
On the other hand, the possibility of a strange star gives a simple 
and unifying  picture for all the systems mentioned above. 
Finally, strange stars have also been speculated to model $\gamma$-ray 
bursters (Haensel \etal\ 1991, Bombaci \& Datta 2000)
and soft $\gamma$-ray repeaters (Cheng \& Dai 1998a).

\vskip 0.6cm
\leftline{{\bf Acknowledgements}}
\vskip 0.2cm
\noindent
I thank my colleagues J. Dey, M. Dey, E.P.J. van den Heuvel, X.D. Li, 
and S. Ray with whom the ideas presented in this talk were developed.   
I am grateful to the Organizing Committee of the Pacific Rim Conference 
on Stellar Astrophysics for inviting me and for financial support. 
Particularly, I thank Prof. K.S. Cheng for the warm hospitality, 
and for many stimulating discussions during the conference.    
It is a pleasure to acknowledge fruitful and stimulating discussions with 
Prof. G. Ripka during the workshop Quark Condensates in Nuclear Matter, 
held at the ECT* in Trento.   

\vskip 0.6cm
\leftline{{\bf In memory of Bhaskar Datta}}
\vskip 0.2cm
\noindent
I dedicate this paper to my great friend and colleague Bhaskar Datta,    
who passed away on december $3^{rd}$ 1999 in Bangalore.  

\newpage


\end{document}